# Gapped Nearly Free-Standing Graphene on an SiC(0001) Substrate Induced by Manganese Atoms

Jinwoong Hwang[a], Ji-Eun Lee[a], Minhee Kang[a], Byeong-Gyu Park[b], Jonathan Denlinger[c], Sung-Kwan Mo[c], and Choongyu Hwang[a],*

[a]Department of Physics, Pusan National University, Busan 46241, Korea

[b]Pohang Accelerator Laboratory, Pohang University of Science and Technology, Pohang 37673, Korea

[c]Advanced Light Source, Lawrence Berkeley National Laboratory, Berkeley, California 94720, USA

**Abstract**

The electron band structure of manganese-adsorbed graphene on an SiC(0001) substrate has been studied using angle-resolved photoemission spectroscopy. Upon introducing manganese atoms, the conduction band of graphene, that is observed in pristine graphene indicating intrinsic electron-doping by the substrate, completely disappears and the valence band maximum is observed at 0.4 eV below Fermi energy. At the same time, the slope of the valence band decreases by the presence of manganese atoms, approaching the electron band structure calculated using the local density approximation method. The former provides experimental evidence of the formation of nearly free-standing graphene on an SiC substrate, concomitant with a metal-to-insulator transition. The latter suggests that its electronic correlations are efficiently screened, suggesting that the dielectric property of the substrate is modified by manganese atoms and indicating that electronic correlations in grpahene can also be tuned by foreign atoms. These results pave the way for promising device applications using graphene that is semiconducting and charge neutral.

*Corresponding author.

Email: ckhwang@pusan.ac.kr

## I. Introduction

Graphene, consisting of a single atomic layer of carbon atoms arranged in a honeycomb lattice, has attracted a lot of research interest due to its two-dimensional nature of Dirac particles realized in a condensed matter system [1,2]. Moreover, graphene is expected to have great potential for fabrication of electronic devices with unique functionalities [3,4]. However, in practice, its gapless semi-metallic nature [5] and inevitable substrate effects, such as hybridization and charge transfer [6–9], make it difficult to utilize graphene as a building block for device applications while preserving the characteristic properties of free-standing graphene. Therefore, decoupling of graphene from the substrate, especially aligning Dirac energy, $E_D$, with Fermi energy, $E_F$, and opening of an energy gap at $E_D$ have been two of the key objectives in graphene research [9–21].

The pursuit for decoupled and gapped graphene has led to exploring the introduction of foreign atoms. Intercalation of foreign atoms breaks the bonding between graphene and the substrate, or compensates for the work function difference between them, leading to transfer of electrons to the substrate that shifts $E_D$ towards $E_F$ [9–15]. The presence of a substrate can lift the sublattice symmetry of graphene, opening an energy gap at $E_D$ [16–18]. Intercalation of foreign atoms further breaks sublattice symmetry, leading to enlargement of the energy gap [19–21]. However, the alignment of $E_D$ with $E_F$ in conjunction with the opening of an energy gap at $E_D$ has been rarely reported so far.

Graphene epitaxially grown on an SiC substrate is a promising candidate for this purpose. The energy gap at $E_D$ is manipulated by the amount of foreign atoms [19–21]. While $E_D$ lies far below $E_F$, graphene gains in metallic properties. On the other hand, the interfacial layer of the SiC substrate is significantly modified by chemical bonding between foreign atoms and silicon atoms in the substrate, such that electrons are transferred back to the substrate, resulting in $E_D$ alignment with $E_F$ [13–15].

In this study, angle-resolved photoemission spectroscopy (ARPES) has been utilized to investigate graphene on an SiC substrate. The presence of manganese atoms in the system not only shifts $E_D$ to $E_F$, but also leads to a metal-to-insulator transition, resulting in gapped nearly free-standing graphene on the SiC substrate. In addition, the observed graphene π band is in agreement with the calculated graphene π band for the case when electron-electron interaction is efficiently screened, indicating the possibility that manganese atoms form a silicide layer with silicon atoms of the SiC substrate, making the interfacial layer of the substrate metallic. These findings suggest that the introduction of transition metals in graphene grown on an SiC substrate provides a viable approach towards the realization of graphene-based electronic devices preserving the characteristic properties of free-standing graphene.

## II. Experimental details

Graphene samples were epitaxially grown on a 6$H$-SiC(0001) substrate under silicon flux in a high vacuum chamber with a base pressure of $3\times10^{-8}$ Torr. Manganese atoms were thermally evaporated by resistively heating a tungsten coil wrapped around a small piece of high-purity manganese (99.9%), followed by another resistive heating of the graphene sample at 600 °C in order to promote intercalations of the deposited metal atoms [22]. Prior to ARPES measurements, the sample was heated to 700 °C to remove air contaminants in a sample preparation chamber with a base pressure of $1\times10^{-10}$ Torr connected to the ARPES chamber. ARPES measurements were performed at Beamline 4A1 of the Pohang Acceleration Laboratory, and Beamlines 4.0.3 and 10.0.1 of the Advanced Light Source, Lawrence Berkeley National Laboratory. ARPES data were taken at 100 K with the energy and angular resolutions of 40 meV and 0.1°, respectively.

## III. Results

Figures 1(a) and 1(b) show ARPES intensity maps of as-grown graphene taken parallel and perpendicular to the ΓK direction of the graphene unit cell, respectively, as denoted by red lines in the insets. The ARPES intensity maps show the characteristic conical energy–momentum dispersion of graphene with two almost parallel dispersions of strong and weak spectral intensity denoted by blue and red arrows, respectively. The pseudospin of quasiparticles in graphene results in a photoemission matrix element effect in which photoelectron intensity is strongly suppressed along the ΓK direction [23], such that one of the branches of the conical dispersion is not observed as shown in Fig. 1(a). The strong and weak spectral intensities correspond to single- and double-layer graphene, respectively. Since photoelectron intensity is closely related to the scattering area of a sample, the strong spectral intensity indicates that single-layer graphene is dominant in the sample. These spectral features are also illustrated in the energy distribution curves (EDCs) shown in Figs. 1(e) and 1(f).

In the measured ARPES intensity map for as-grown graphene, $E_D$, denoted by a black arrow, is observed at 0.4 eV below $E_F$, indicating intrinsic electron doping in the presence of the SiC substrate [24]. One can notice that the upper conical dispersion, the so-called $\pi^*$ band or conduction band, and the lower conical dispersion, the so-called $\pi$ band or valence band, do not merge at a single point. The merging point is elongated along the energy direction, leaving a high spectral intensity around $E_D$ [25], inconsistent with the zero density of states of ideal graphene [5]. A carbidic layer, the so-called buffer layer, exists in between graphene and the SiC substrate [26], and exhibits the same geometric structure as graphene in the absence of the conical dispersion due to the formation of covalent bonds with the substrate. The presence of the buffer layer breaks sublattice symmetry in single-layer graphene, leading to the opening of an energy gap at $E_D$ [16,18]. The buffer layer also leads to an in-gap state that is observed only for single-layer graphene on an SiC substrate [18,25]. These two features are responsible for the elongated spectrum at $E_D$ observed in Fig. 1(b).

The introduction of manganese atoms into the system makes three notable changes in the electron band structure of graphene as shown in Figs. 1(c) and 1(d). A non-dispersive band emerges at 2.8 eV below $E_F$ as denoted by the black dashed line. It is not observed in as-grown graphene shown in Figs. 1(a) and 1(b), but bears similarity to the localized $d$ band observed in graphene with transition metals such as nickel, cobalt, and iron [7,9,11]. The non-dispersive band is well resolved in the EDCs shown in Figs. 1(g) and 1(h), confirming the presence of manganese atoms in the system. Moreover, it is also important to note that the conduction band observed in as-grown graphene completely disappears, indicating that charge neutrality of free-standing graphene is recovered although graphene is standing on the SiC substrate. Finally, the valence band maximum is observed at ~0.4 eV below $E_F$ as shown in Fig. 1(d) and denoted by the blue dashed-line in Fig. 1(h), while the spectral intensity between the valence band maximum and $E_F$ is strongly suppressed. When the difference between the valence band maximum and $E_F$ is typically half the energy gap, the energy gap of graphene on an SiC substrate with manganese atoms can be as much as ~0.8 eV. In other words, the charge carrier density of ~$10^{13}$ cm$^{-2}$ of as-grown graphene is significantly reduced, so that metallic graphene becomes a band insulator. In epitaxial graphene on an SiC substrate, the presence of the substrate breaks the sublattice symmetry of graphene, such that the conduction and valence bands of graphene exhibit a finite separation of ≤ 0.26 eV [18]. The larger energy gap observed in this system indicates that the sublattice symmetry of graphene might be further lifted by the presence of manganese atoms.

### IV. Discussion

The electronic structures observed in graphene on an SiC substrate without and with manganese atoms and their geometric structures are summarized in Fig. 2. The metal-to-insulator transition concomitant with the shift of $E_D$ to $E_F$ bears two possibilities. First, manganese atoms can induce hole doping in graphene, compensating electron doping by the substrate. However, a previous scanning tunneling spectroscopy study shows that manganese atoms induce electron doping [22], excluding this possibility. Epitaxial graphene on an SiC(0001) substrate is electron-doped due to the formation of a Schottky barrier in between graphene and the substrate [24]. The Schottky barrier can be lowered, transferring charge back to the substrate, when manganese atoms modify the geometric structure of the interfacial layer of the substrate. Indeed, when gold [13] or tin [15] atoms are introduced in graphene on an SiC substrate, these atoms chemically bond with silicon atoms in the SiC substrate, recovering charge neutrality of the overlying graphene.

Detailed analysis of the electron band structure of graphene supports that the substrate is modified by the presence of manganese atoms. Figure 3(a) shows the graphene π band of as-grown graphene (red curves) and graphene with manganese atoms (blue curves) obtained using a Lorentzian fit to the momentum distribution curves (MDCs) for the ARPES intensity maps shown in Figs. 1(a) and 1(c), respectively. To study the change in the dispersion induced by manganese atoms, the dispersions are shown with respect to $E_D$ and compared to the calculated π band (black curve) using the local density approximation (LDA) method [27]. While the slope of the dispersion at $E_D$ decreases from $0.89 \times 10^6$ m/s for as-grown graphene to $0.86 \times 10^6$ m/s for graphene with manganese atoms, the latter value is similar to the value for the LDA band of $0.85 \times 10^6$ m/s. More importantly, the overall slope of the dispersion of graphene with manganese atoms is lower than that of as-grown graphene, approaching the value for the LDA band.

The slope of the graphene π band is strongly influenced by dielectric environment [28]. When graphene is placed on a metallic substrate, electron–electron interaction is screened by the metallic background from the substrate. Such screening results in the graphene π band convergence towards the LDA band, which describes the case of efficiently screened electron-electron interaction [6,29]. The energy–momentum dispersion of graphene with manganese atoms is in good agreement with the LDA band as shown in Fig. 3(a). This agreement indicates the possibility that the interface of SiC underneath graphene becomes metallic, leading to enhanced screening of electron-electron interaction in graphene compared to that for as-grown graphene. A plausible origin of this screening is the formation of silicon-manganese bonds, i.e., manganese silicide. Such silicide may not be well ordered, possibly resulting in random site potentials, which is exhibited as the enlarged MDC width shown in Fig. 3(b) (the peak observed at ~0.4 eV below $E_F$ and at $E_F$ denoted by a black arrow is due to the absence of a well defined quasiparticle state by the energy gap at $E_D$ [25]). The formation of a disordered silicon-metal layer in the SiC substrate has been reported to significantly reduce substrate influence for the cases of epitaxial graphene with gold [13] and tin [15]. The silicon-metal layer could be responsible for the charge transfer back to the substrate, which realizes charge-neutral graphene and further lifting of the sublattice symmetry, leading to a larger energy gap than that of as-grown graphene, as observed in Fig. 1 and which is schematically shown in Fig. 2. However, it is important to note that the understanding of the role of manganese in the formation of gapped nearly free-standing graphene requires additional experimental and theoretical studies. X-ray photoemission spectroscopy will be able to confirm the formation of the silicide layer and model calculations can be conducted to simulate the enhancement of the broken sublattice symmetry.

The interaction of graphene with foreign atoms can induce an interesting spin state in the system. The hybridization between the graphene π band and the cobalt 3*d* or platinum 5*d* state results in polarized spin states in graphene [7,30]. In addition, the graphene π band participates in the formation of a spin-dependent many-body ground state in the presence of the cerium 4*f* state [31]. In fact, when manganese atoms form a cluster, their interatomic spin coupling leads to the RKKY (Ruderman-Kittel-Kasuya-Yosida) type interaction in graphene [32]. Furthermore, the formation of silicide can be boosted in the case of manganese atoms with high chemical reactivity with silicon [33] by performing annealing, that was applied to the samples during the cleaning process before the ARPES measurements. In this way, spin interactions are disturbed, resulting in the absence of magnetic effects. On the other hand, the spontaneous polarization by charge redistribution from silicide [15,34] is assumed to be responsible for the decoupling of graphene from the substrate and the metal-to-insulator transition.

## V. Conclusions

Electronic properties of graphene on an SiC(0001) substrate in the presence of manganese atoms have been investigated using the ARPES technique. The introduction of manganese atoms in the system allows graphene to approach its charge neutrality and to undergo a metal-to-insulator transition. In addition, the electron–electron interaction in graphene is sufficiently screened, which is confirmed by the graphene π band agreement with the LDA band, suggesting the possibility that manganese atoms form a metallic silicide layer on top of the SiC substrate. These findings provide not only the experimental evidence of gapped nearly free-standing graphene on an SiC substrate, but also a promising route towards the application of graphene, which is semiconducting and charge neutral, to electronic devices.

**Acknowledgements**

This work was supported by a 2-Year Research Grant from the Pusan National University.

**Figure Captions**

Fig. 1. (a–d) ARPES intensity maps of as-grown graphene (a,b) and graphene with manganese atoms (c,d) taken along and perpendicular to the ΓK direction around the K point as denoted by red lines in the inset. The blue and red arrows denote the electron band structure of single-layer and double-layer graphene, respectively. The black arrow denotes the Dirac energy of single-layer graphene. (e–f) Energy distribution curves for the ARPES data shown in panels (a–d). The black dashed line denotes the manganese 3*d* state that is non-dispersive. The blue dashed-line in panel (h) denotes the π band of graphene with manganese atoms.

Fig. 2. Schematic of the geometric and electronic structure of graphene before (a) and after (b) the introduction of manganese atoms in the system. Red and blue spheres denote carbon sublattices.

Fig. 3. (a) Energy–momentum dispersions of as-grown graphene (red curve) and graphene with manganese atoms (blue curve) along the ΓK direction of the graphene unit cell denoted by a red line in the inset, compared to the calculated dispersion using the LDA method (black curve). The dispersions are shown with respect to $E_D$. (b) The full width at half maximum of the momentum distribution curves of as-grown graphene (red dots) and graphene with manganese atoms (blue dots).

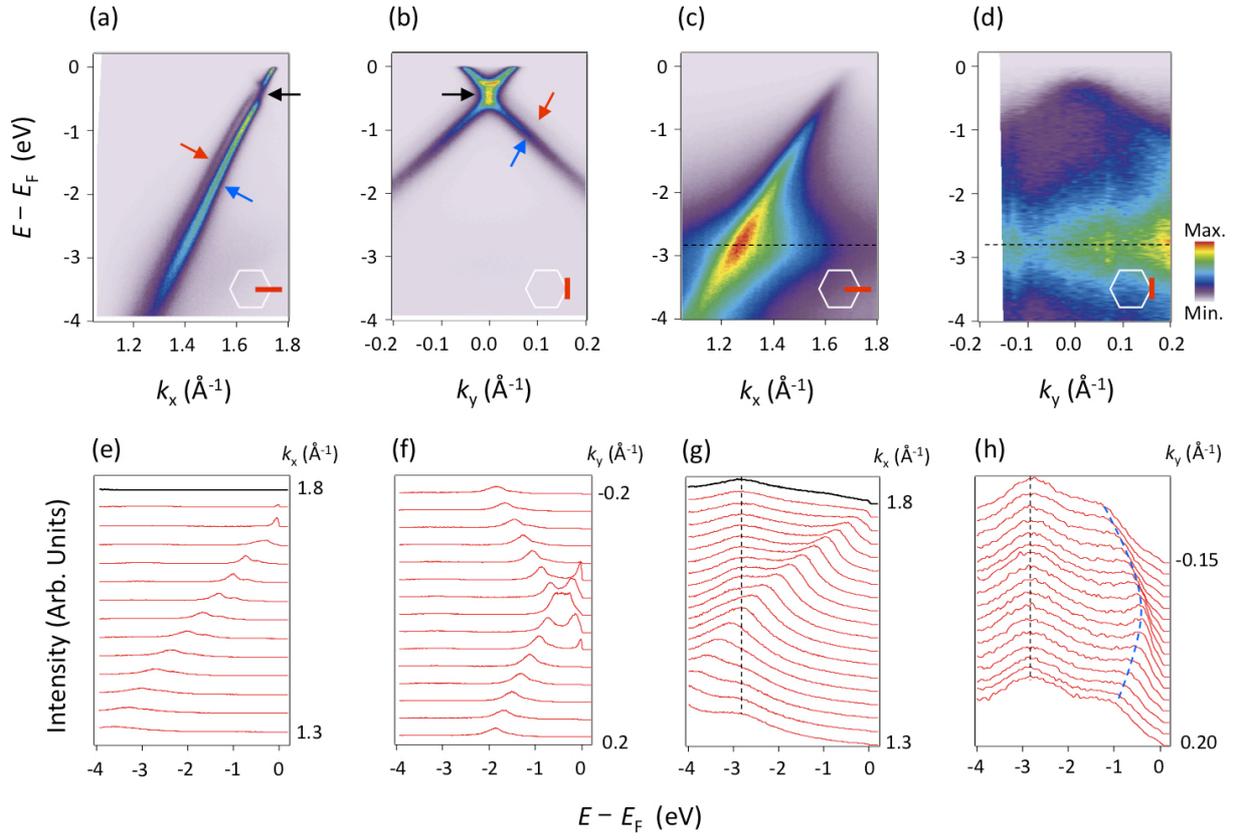

FIG. 1.

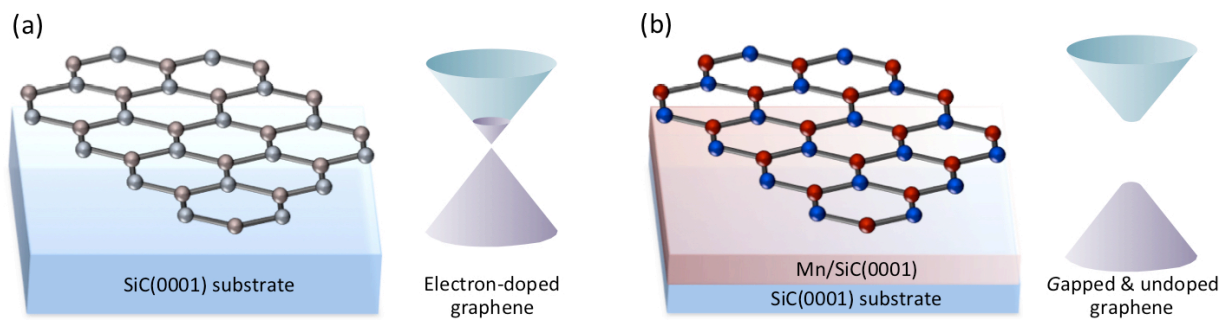

FIG. 2

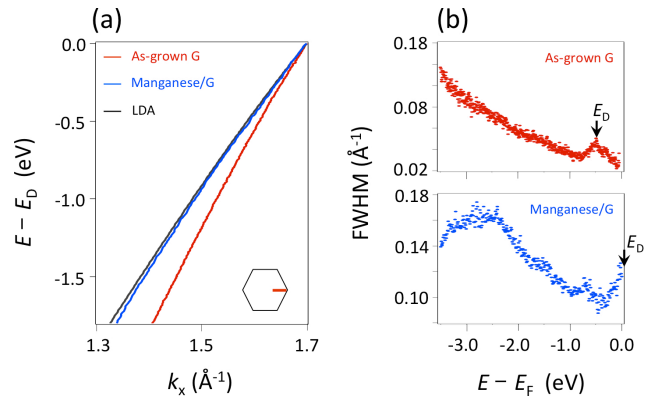

FIG. 3